\documentstyle[12pt]{article}

\def\bea{\begin{eqnarray}}
\def\eea{\end{eqnarray}}

\begin{document}

%
%
 
%
\renewcommand{\topfraction}{0.99}
\renewcommand{\bottomfraction}{0.99}
  
\title
{\Large The Promise of String Cosmology \footnote{This article was commissioned in conjunction of the Workshop on String Cosmology held from July 24 - Aug. 4 2000 at PIMS, Univ. of British Columbia and is published in APCTP Bulletin \# 6, (2001). Brown preprint BROWN-HET-1243.}}
\author{Robert H. Brandenberger,\\ 
Department of Physics, Brown University,  
Providence, RI 02912, USA.}
\date{\today} 
\maketitle
\begin{abstract}
The interplay between string theory and cosmology
is very promising. Since string theory will yield
a quantum theory of space-time and unify all forces of
nature, it has the potential
of addressing many of the conceptual problems of the
current models of early Universe cosmology. In turn, 
cosmology is the most obvious testing ground in the
effort to construct non-perturbative string theory,
and can provide the crucial connection between
theory and experiment/observation \footnote{For a more
technical overview of the existing approaches to connecting
string theory and cosmology, the reader is referred in
\cite{Easson}.}.
\end{abstract}

\vskip 0.8cm

{\bf Early Universe cosmology} has made spectacular progress over the
past two decades, driven both by a wealth of new observational
data and by theoretical breakthroughs. In particular, a
scenario for the very early Universe, the {\bf inflationary
Universe} scenario \cite{Guth}, has emerged which has led to solutions
of some of the deep mysteries of standard big bang cosmology,
and which has made some quite specific predictions for the
geometry of the Universe and for the spectrum of cosmic
microwave background (CMB) anisotropies which have been confirmed
by the most recent observations \cite{Boomerang,Maxima}.

In spite of its success, the inflationary Universe scenario
suffers from some deep conceptual problems \cite{RB99} which have
prevented it from achieving the status of an honest physical
theory. For example, scalar-field-driven inflationary
cosmology is incomplete in the same sense as the standard
big-bang cosmology was: in the context of inflationary
cosmology an initial singularity is inevitable, thus
making it impossible to formulate a consistent initial
value problem. More seriously, the successful predictions
of inflationary cosmology for the spectrum of the CMB
depend - in the current models of inflation in which the
exponential expansion of space is driven by the potential
energy of a scalar matter field - on extrapolating the
physical theory into a regime where its foundations (namely
classical general relativity and weakly coupled quantum field
theory) are known to break down \footnote{This is the so-called
trans-Planckian problem for inflationary cosmology \cite{BM00}.}. The Achilles
heel of inflationary cosmology is the cosmological constant
problem. How do we know that the unknown mechanism which cancels
the vacuum energy today does not also cancel the transient 
vacuum energy which is required to drive inflation? Quite apart
from these conceptual problems, a convincing realization of
inflation in the context of our current models of particle
physics based on quantum field theory is lacking.

Recent observations are also beginning to challenge some of
the premises of our current cosmological models. These
observations point to the existence of a dark energy
component in the Universe which makes up the bulk of the
energy density but which does not cluster \cite{SN1,SN2}. Furthermore,
recent
measurements and numerical simulations show a mismatch
between the predicted and observed galaxy halo structures
which may imply that the dark matter is not cold (see e.g. \cite{SS} and
references therein).

{\bf String (or M-) theory} is the best candidate for a
unified theory of all forces and as such for a quantum theory
of space-time and matter. Thus, it will resolve
the conflict between quantum field theory and
general relativity, and therefore it has the potential of 
addressing many of the open issues in early Universe cosmology.
If successful in this challenge, string theory will have
made an important contact with
experiment/observation, and will no longer be able
to be criticized for lacking a connection to data and for thus being pure
mathematics rather than a physical science.
The successful determination of the entropy of an extremal black hole
in the context of string theory \cite{SV} is a first major success in
establishing a link between string theory and cosmology.

Until recently there has not been much work on the interface between
string theory and cosmology. 
However, it has been clear since the late 1980's 
that string theory has the potential to solve some of the problems of standard
cosmology. For example, it was argued \cite{BV89}, in the context of perturbative string theory,
and assuming that all spatial dimensions are toroidally compactified,
that as a consequence of t-duality the physics at very small radii is equivalent to that
at very large radii, hence eliminating the cosmological singularities. Moreover,
string winding modes may yield a mechanism which allows at most three spatial dimensions
to become large, thus providing a dynamical resolution of the potential
embarrassment that superstring theory is consistent only in nine (or ten in the
context of M-theory) spatial dimensions. This example demonstrates that with string
theory, questions about the Universe can now be addressed scientifically which before
the advent of string theory were exclusively in the realm of philosophy.

In the past five years string theory has undergone a ``second revolution". With the
discovery of {\bf D-branes} \cite{Pol} it has
become clear that there are many more fundamental degrees of freedom than just strings.
Dualities between the previously known five consistent string theories were discovered
which indicate that there is a common underlying theory, {\bf M-theory}. The different
string theories (and also 11-d supergravity) correspond to certain corners of moduli space of
M-theory. A further new
important tool is the {\bf AdS-CFT correspondence} \cite{Malda} which relates classical
gravity on
an anti-de-Sitter background to a conformal field theory living on the boundary.
Recently, much attention has been focused on {\bf brane world scenarios}, scenarios
based on a combination of the above ideas in which the matter fields of our Universe are
confined to a three-brane and only gravity lives in the higher-dimensional bulk (see
e.g. \cite{HW}). In
general, the spatial dimensions transverse to the branes must be very small in order to
avoid too large deviations from Newton's gravitational laws. An interesting
suggestion first put forward without a direct connection to string theory \cite{ADD} \footnote{See
\cite{AADD} for a realization of this idea in the context of string theory.}
is that two of the extra dimensions are large (mm scale). Since the four dimensional
effective gravitational constant is related to the fundamental higher dimensional one via the size
of the extra dimensions, the fundamental gravitational constant can be reduced to the TeV
scale, provided the large extra dimensions have a size of the order of 1mm. This provides the
potential for solving the mass hierarchy problem of elementary particle physics. By
constructing
models in which gravity remains confined to the brane, it is possible to make the
extra dimensions large (in fact even infinite \cite{RS2}). 

On the other hand, there is still no non-perturbative formulation of M-theory; not even its
fundamental physical degrees of freedom are known. In at least one approach to M-theory \cite{BFSS},
the underlying theory is postulated to be a quantum mechanical matrix model. If this is true,
then some of the non-commutativity of the matrix model will be reflected in non-commutativity of the
space-time coordinates at a microscopic level.
Returning to the more traditional description of M-theory, another
serious problem results from the fact that the moduli
space of vacuum states corresponding to any of the string theories is huge, and 
many of the predictions
for particle physics as well as for cosmology depend on which vacuum state is chosen. A key
challenge to string cosmology is to find a mechanism which distinguishes the vacuum state
corresponding to our Universe. Since the moduli correspond to massless fields,
new problems for cosmology appear: why don't we see the massless particles associated
with the moduli fields, and if this question is answered in the usual way by
invoking some non-perturbative mass generation for the moduli, then why don't the
moduli over-close the Universe \cite{Dine}? 

However, even without
first having to solve these difficult problems, there
are interesting avenues for string cosmology. A particularly interesting issue is
to determine whether there are string-specific ways to obtain a period of inflationary expansion in
the early Universe. If the answer to this question is affirmative, then it is of utmost
importance to try to find observational signatures which could distinguish these
string-specific inflationary models from scalar-field-driven inflation. The answer to
this question will inevitably require looking beyond the power spectra of density
fluctuations and of CMB anisotropies, since any set of power spectra can be obtained
by suitable choice of a toy model scalar field potential. It may, however, also turn
out that string cosmology does not yield anything that looks like the usual inflationary
scenario, but that the resulting theory nevertheless provides an alternative to inflation in solving
the problems of standard big bang cosmology such as the horizon and flatness problems. This
possibility is realized in the pre-big-bang approach to string cosmology \cite{GV} based
on dilaton gravity as the low-energy effective theory consistent with the symmetries of
string theory. In this scenario the background in the Einstein frame is in fact
contracting super-exponentially in the pre-big-bang phase, leaving an effective Hubble
constant whose absolute value is increasing in time. In pre-big-bang cosmology the spectrum
of scalar metric fluctuations stemming from the dilaton gravity sector is not scale-invariant
as it would be in inflationary cosmology \cite{GVM}. A more radical alternative to inflation
which can solve the homogeneity and flatness problems is the varying-speed-of-light
scenario in which the speed of light in the early stages of the evolution of the Universe
is postulated to be many orders of magnitude larger than at the present time
\cite{Moffat,AM}.

Another direct observational issue which string cosmology should address concerns the origin of dark
energy. Since string theory leads to a large number of scalar moduli fields, it is of great
interest to explore the potential of these fields for explaining the origin of dark energy.
Possibly related to this issue is the question of the smallness of the observed cosmological
constant. Finally, even without knowing the exact vacuum state of string theory it should be
possible to study whether string theory will indeed wash out the singularities of standard
and inflationary cosmology.

Interest in string cosmology is growing. From July 24 - August 4, 2000, about 45
physicist working on string theory, non-perturbative field theory and cosmology
gathered at the University of British Columbia (UBC) in Vancouver
for a workshop devoted to string cosmology. This workshop was co-sponsored by the Pacific
Institute for Mathematical Sciences (PIMS), the Canadian Institute for Advanced Research (CIAR)
and by the APCTP. The invited participants included B. Greene (Columbia Univ.), N. Kaloper
(Stanford), L. Kofman (CITA, Univ. of Toronto), D. Lowe (Brown Univ.), B. Ovrut (Univ. of
Pennsylvania), S. Ramgoolam (Brown Univ.), S. Sin (Hanyang Univ.), D. Son (Columbia
Univ.), P. Steinhardt (Princeton Univ.),
H. Verlinde (Princeton Univ.), G. Veneziano (CERN) and A. Zhitnitsky (UBC) \footnote{See the web site
http://kepler.physics.ubc.ca/\~pfs99 for details about the workshop.}.
Many different approaches to string cosmology were represented at this workshop. In the
following discussion of various approaches to string cosmology I will limit myself to those
which were discussed extensively at the workshop. 

Based on the conclusions of one of the most conservative approach to string
cosmology, namely {\bf Pre-Big-Bang Cosmology} \cite{GV}, it is clear that we should
expect important differences in the evolution of the early Universe in string cosmology
compared to scalar-field-driven inflationary cosmology. Pre-big-bang cosmology is a low
energy description of physics which takes into account all of the low energy modes in
string theory, i.e. not only the graviton but also the dilaton and the anti-symmetric
tensor field (the latter, however, does not play an important role for the background
dynamics at the level of the homogeneous equations of motion \footnote{As was mentioned at the workshop, a non-vanishing tensor field decreases the set of inhomogeneous initial conditions which can lead to the pre-big-bang evolution.}). The symmetries and equations of motion for dilaton gravity give rise to a scenario
of cosmology in which the Universe starts out in a state near the perturbative string
vacuum, goes through a dilaton-driven phase of contraction during which the Hubble radius
$H^{-1}(t)$ ($H$ being the expansion rate of the Universe and $t$ denoting time) shrinks
faster than the physical length of fixed comoving scales, thus simultaneously solving
the horizon problem of standard cosmology and giving rise to a mechanism for
structure formation similar as in the usual inflationary cosmologies: quantum vacuum fluctuations
existing during the contraction phase on sub-Hubble lengths get frozen in when the Hubble
radius becomes smaller than the wavelength of the fluctuation, they are amplified
super-adiabatically while the wavelength is larger than the Hubble radius, and then re-enter
the Hubble radius in the late Universe as classical density fluctuations. As the Universe
contracts, the curvature increases. Eventually, in a way which unfortunately cannot be
described by the equations of dilaton gravity, the Universe makes a transition to an
expanding Friedmann cosmology related to the contracting phase via a duality transformation.
Note that in pre-big-bang cosmology, both the background evolution and the spectrum of
induced fluctuations is different from that of scalar-field-driven inflationary models.
The density fluctuations from the dilaton sector are not scale invariant. In fact, they
are completely unimportant on large length scales. The other fields present in 
pre-big-bang cosmology (such as the axion) will, however, generate fluctuations which can be
scale-invariant \cite{Easther}. Since these fluctuations start out as isocurvature
fluctuations, they lead to different predictions for observables such as the spectrum of
CMB anisotropies than the scalar metric fluctuations of inflationary cosmology. The lesson
we learn is that it is quite likely that the predictions of string cosmology will differ
enough from those of standard inflationary cosmology to enable observations to probe the
physics of the Planck scale.

Pre-big-bang cosmology suffers from the graceful exit problem: in the context of the equations
of motion of dilaton gravity, the contracting branch of the evolution terminates in a singularity,
and the expanding branch emerges from an equivalent singularity. Since the singularity occurs
in the high curvature regime, it is clear that the effective dilaton gravity action no longer
describes the stringy physics in a correct way close to the singularity. A challenge for
string cosmology is to understand if true stringy physics can solve the graceful exit
problem and lead to a smooth transition from the contracting dilaton-dominated phase to the
expanding phase with fixed value of the dilaton. Although there have been several attempts to
go beyond dilaton gravity in order to solve the graceful exit problem (see e.g. \cite{Ven} for
a recent overview), there is not yet a convincing approach. An important open problem for
string cosmology is to develop a truly stringy, non-perturbative version of the pre-big-bang
scenario, or to find out how the scenario must be modified in order to become a truly
stringy cosmological model.

Another promising approach to string cosmology is based on Horava-Witten theory \cite{HW} and is
sometimes called {\bf heterotic M-theory cosmology} \cite{Ovrut}. The theory is based on one
particular ray in the moduli space of M-theory, namely the compactification of 11-dimensional
supergravity on an orbifold $S^1 / Z_2$. This leads to two distinguished co-dimension one planes,
the orbifold fixed planes. As usual, it is assumed 
that, in addition, six spatial dimensions are compactified on a special type of Calabi-Yau 
manifold, and that there is a set of $E_8$ gauge supermultiplets on each orbifold plane.
The reason for adopting this particular orbifold compactification 
is that it allows for chiral fermions \cite{HW}. The resulting
theory is dual to the strongly coupled $E_8 \times E_8$ heterotic string theory. 
Matching at tree level to the phenomenological gravitational and grand unified gauge couplings,
one finds that the orbifold must be larger than the Calabi-Yau radius. This suggests
that there is a regime in which the dynamics of the Universe is effectively five-dimensional.
The resulting five-dimensional effective theory is a gauged $N = 1$ supergravity
coupled to four-dimensional boundary gauge and matter supermultiplets (see e.g. \cite{Ovrut2} for
a recent survey). Initially, only the effective four-dimensional equations were studied. However,
it has recently been realized that the dynamics in the orbifold direction is quite nontrivial
and may lead to new effects of great interest to cosmology. In particular, the equations of
motion in five dimensions admit as a static solution a pair of three-branes parallel to the orbifold
fixed planes. It is tempting to try to obtain inflation from the nontrivial dynamics
involving the orbifold direction. Note, however, that as in pre-big-bang cosmology, many truly
stringy effects are frozen out in the effective action-based analysis of heterotic M-theory
cosmology.  

A rather different approach to string cosmology is based on proceeding in analogy to the usual
starting point of big bang cosmology, which is the assumption that the Universe starts out with
Planck size and emerges from a
hot initial state in which all degrees of freedom were highly excited. It seems quite
natural to generalize this starting point to 
string cosmology and to assume \cite{BV89} that the Universe starts out with all spatial
dimensions of string scale and all fundamental degrees of freedom excited according to their
thermal distribution. In particular, this means that the winding modes of strings and branes are
highly excited. This approach is called the {\bf brane gas approach} to string cosmology. In this
approach, all spatial dimensions start out compact and the question for cosmology is not why
certain dimensions are compactified, but why a fixed number of dimensions can dynamically
decompactify (in the sense of their radii becoming larger than the present Hubble radius of
the Universe). There have been some interesting recent developments \cite{MR,Sin} 
along the lines of this approach to string cosmology. In particular, assuming that
all spatial dimensions are compactified on a torus, it has been shown that
whereas at early times (small sizes) the higher dimensional branes dominate the evolution, it
is the fundamental strings which have the most important effect at late times \cite{abe}. Thus, the mechanism
of \cite{BV89}, which only allows three spatial dimensions to become large, survives in the modern
view of string theory. In addition, the duality under inversion of the radius of the torus is maintained,
thus assuring that no physical cosmological singularities will arise. It is important to generalize
these considerations to more realistic background spaces.

The approach to string cosmology which at the present time is attracting most attention is the class
of {\bf brane-world scenarios} (see e.g. \cite{Kaloper} for work along these lines
presented at the APCTP/CIAR/PIMS string cosmology workshop). The starting 
point of this approach is the assumption that our
matter fields are confined to a three-brane (four space-time dimensions) embedded in a higher-dimensional
(usually five-dimensional) bulk space-time. An important realization is that it is possible 
\cite{RS2} to
confine gravity on one of the branes by having the four-dimensional metric depend on the fifth
coordinate (the orbifold coordinate in the language of heterotic M-theory cosmology). In many
cases, the bulk metric is  AdS, and this gives rise to a tantalizing connection with the
AdS-CFT correspondence. However, most 
work along these lines is not directly justified from string theory. 
There appear to be an unlimited number of versions of the basic scenario. As was stressed at the
APCTP/CIAR/PIMS string cosmology workshop, an urgent task for string cosmology is to derive the
basic rules which a brane-world scenario must obey if it is to emerge from string theory. Is the
brane an orbifold fixed plane (as it is in heterotic M-theory cosmology) or is it a D-brane? How
many branes are there in the scenario? Are there matter fields in the bulk? At the present time,
most work on brane-world cosmology is based on studying the background gravitational equations
of motion of the brane in the presence of the bulk, making use of the general relativistic Israel
matching conditions. It has been realized that the equations for the dynamics of our brane in the brane-world scenario
may be quite different from the usual four-dimensional general relativistic equations. This provides
stringent constraints on brane world cosmologies. However, it also provides the chance to obtain new
effects, or to provide new sources of inflation. In order to make contact with the wealth of
recent cosmological observations on the distribution of light and CMB light, it is crucial to
extend the theory of cosmological perturbations, the basis for making such a contact, from the
usual four-dimensional setting to the brane-world scenarios. Initial work along these lines was
presented at the workshop (see \cite{Dorca} and references therein). A limitation which the
brane-world scenarios share with some of the other approaches to string cosmology is that most of
the analysis is based on a truncation of the dynamics to low energy modes, thus eliminating many
truly stringy effects. It is, however, possible to obtain interesting new effects for cosmology.
For example, if our Universe is a three-brane moving in a bulk black hole background, it is
possible \cite{Stephon} to find a realization of the varying light scenario for solving the
problems of standard cosmology. 

Ultimately, however, a complete cosmological model should be based on the non-perturbative formulation
of M-theory. Starting from the matrix theory approach to M-theory (see e.g. \cite{BFSS,DVV}), it
seems likely that the resulting space-time theory will be non-commutative at the string scale (see
e.g. \cite{ramgosk} for work along these lines presented at the workshop). One limitation of the
matrix theory approach to M-theory \cite{HV} is that the theory is at the present time only defined
in light-cone coordinates, which makes the connection to a space-time formulation very challenging.

A generic problem which appears in all approaches to string cosmology is the moduli problem. What
freezes the value of the dilaton and what sets the scale of the extra dimensions? Usually it
is assumed that some non-perturbative physics is responsible for the answers. If an effective
potential provides the required mechanism (see e.g. \cite{Huey} for some recent work presented in
Vancouver), then why does the value of the potential at its minimum not lead to another source
for the cosmological constant? Thus, the moduli problem and the cosmological constant problem
appear to be related. There are some new approaches to solving the cosmological constant problem
based on the brane-world scenario, e.g. a model by Verlinde \cite{HV2} in the context
of a theory with two branes, the physical brane and the Planck brane, which assumes that 
the cosmological constant on the Planck brane is protected  by supersymmetry, and that it flows
to a small cosmological constant on the physical brane via renormalization group flow.

The field of string cosmology is still in its infancy. Although many interesting approaches
to connecting string theory and cosmology exist, most of them suffer from our ignorance about
the true non-perturbative formulation of string theory. The promises of the field, however,
are great, as witnessed by the excitement expressed at the APCTP/CIAR/PIMS string cosmology
workshop
 
\vskip 0.6cm
\centerline{\bf Acknowledgements}

I wish to thank the APCTP, CIAR and PIMS for their generous support of the 
{\it String Cosmology Workshop} held at UBC in Vancouver as part of the
{\it Frontiers of Mathematical Physics} workshop series jointly organized
by PIMS and APCTP. I also thank R. MacKenzie for useful comments on
the draft. The author's research was supported in part  by the U.S. 
Department of Energy under Contract DE-FG02-91ER40688, TASK A.

\end{document}